\documentclass[a4paper,11pt]{article}
\pdfoutput=1

\usepackage{jheppub}

\usepackage[T1]{fontenc}
\usepackage[utf8]{inputenc}

\usepackage[english]{babel}
\usepackage{mathtools,amssymb}
\usepackage{braket,slashed}
\usepackage{graphicx}
\usepackage{booktabs,dcolumn,multirow}

\newcommand{\diag}{\mathrm{diag}}
\newcommand{\bs}{\boldsymbol{s}} 
\newcommand{\mX}{\mathcal{X}}
\newcommand{\Qbar}{\mkern 1.5mu\overline{\mkern-1.5mu Q \mkern-1.5mu}\mkern 1.5mu}
\newcommand{\pmatr}[1]{\begin{pmatrix} #1 \end{pmatrix}}

\newcommand{\uV}[2]{\mathcal{V}_{#1}^{#2}}
\newcommand{\cVec}{V}
\newcommand{\cAx}{A}

\newcommand{\eqn}[1]{eq.~(\ref{#1})}
\newcommand{\eqns}[2]{eqs.~(\ref{#1})-(\ref{#2})}
\newcommand{\Eqn}[1]{Eq.~(\ref{#1})}
\newcommand{\Eqns}[2]{Eqs.~(\ref{#1}) and (\ref{#2})}

\title{$U(1)$ flavour symmetries as Peccei-Quinn symmetries}

\author[a]{Fredrik Bj\"orkeroth,}
\author[b]{Luca Di Luzio,}
\author[c]{Federico Mescia,}
\author[a]{Enrico Nardi}

\emailAdd{fredrik.bjorkeroth@lnf.infn.it} 
\emailAdd{luca.diluzio@pi.infn.it}
\emailAdd{mescia@ub.edu}
\emailAdd{enrico.nardi@lnf.infn.it}

\affiliation[a]{INFN, Laboratori Nazionali di Frascati, C.P. 13, 100044 Frascati, Italy}
\affiliation[b]{%
  Dipartimento di Fisica, Universit\`a di Pisa and INFN, Sezione di Pisa, \\ 
  Largo B. Pontecorvo 3, 56127 Pisa, Italy
} 
\affiliation[c]{%
  Departament de F\'isica Qu\`antica i Astrof\'isica, Institut de Ci\`encies del Cosmos (ICCUB), \\
  Universitat de Barcelona, Mart\'i Franqu\`es 1, E08028 Barcelona, Spain
}

\abstract{
We investigate to what extent a generic, generation-dependent $U(1)$
symmetry acting on the quark Yukawa operators can reduce the number
of free parameters by forcing some entries in the Yukawa matrices to
vanish.  The maximal reduction compatible with CP violation yields
nine real parameters and one phase, which matches the number of
physical observables, implying that such models have no free
parameters.  We derive a set of results: (i) the only possible
structures have the form $M_4 \oplus M_5$, where the subscripts
indicate the number of real parameters in the Yukawa matrices,
(ii) there are only two inequivalent Yukawa structures, each one
giving rise to six different models depending on quark flavour
assignments, (iii) the $U(1)$ symmetries that generate these
textures all have a QCD anomaly, and hence are Peccei-Quinn 
symmetries, reinforcing the idea of a possible connection between the
quark flavour puzzle and the axion solution to the strong CP
problem, (iv) in some cases the contributions to the QCD anomaly of
two generations cancels out, and this opens the possibility that the
axion coupling to nucleons could be strongly suppressed.
Flavour-violating axion couplings to quarks are completely fixed, up
to the axion decay constant, providing a non-trivial complementarity
between low-energy flavour-violating processes and standard axion
searches.
}


\begin{document}
\maketitle
\flushbottom

\section{Introduction} 

The origin of flavour remains one of the least understood aspects of
the Standard Model (SM): the large majority of the SM free parameters
are related to flavour and, at least in the quark sector, their values
do not appear to be random.
Quark masses are strongly hierarchical and, when the same ordering is
chosen in the up and down sectors (for example from light to heavy), the
Cabibbo-Kobayashi-Maskawa (CKM) mixing matrix is close to diagonal.
This implies a surprising degree of alignment between the up and down
Yukawa matrices: either the weak and mass eigenstates are nearly aligned 
in both sectors or some mechanism ensures that the misalignments are 
quantitatively similar.

Many efforts have been made to address the flavour puzzle, often by
invoking flavour symmetries. These may be Abelian or non-Abelian,
global or discrete.  In particular, non-Abelian family symmetries have
received much attention by model builders, as they allow for
unification of the fermion generations.  They are often discrete and
are frequently used in conjunction with gauge unification, see
e.g.~\cite{King:2017guk,King:2013eh} for reviews.  Another popular
approach to accommodate quark mass hierarchies and small mixing angles
is to postulate a global $ U(1) $ symmetry (or possibly a discrete
$ \mathbb{Z}_N $ subgroup).  The canonical example is the
Froggatt-Nielsen mechanism \cite{Froggatt:1978nt}, whereby the
symmetry forbids most fermion Yukawa
couplings at the renormalizable level.  The symmetry is spontaneously broken by the vacuum
expectation value (vev) of a SM singlet flavon field, after which a
set of effective operators arises that couples the SM fermions to the
electroweak Higgs boson. The hierarchy of fermion masses results from
the dimensional hierarchy among higher-order operators, which are
suppressed by powers of a dimensionless, and conveniently small,
symmetry-breaking parameter, with the suppression powers determined by
the Abelian charges assigned to the fermion fields.

A different approach, put forth already in the 1970s
\cite{Fritzsch:1977za,Fritzsch:1977vd}, aims to reproduce the data
with a reduced number of free parameters.  While this generally allows
one to compute some of the observables in the Yukawa sector, the more
ambitious goal is to reveal some well-defined structure that could
shed light on the mechanism at the origin of the flavour architecture.
This strategy remained actively pursued during the last decades of the
past
century~\cite{Stech:1983zc,Dimopoulos:1991za,Giudice:1992an,Ramond:1993kv}.
Since then, the viability of Yukawa matrices with a reduced set of
non-vanishing entries and different numbers of textures zeros has been
systematically studied (see
e.g.~\cite{Xing:2015sva,Giraldo:2015ffa,Giraldo:2011ya}).  However, these types of studies in
general do not specify the detailed mechanism (presumably some
symmetry) which enforces the vanishing of specific entries in the
Yukawa matrices.

In this work we study to what extent a generation-dependent
global $U(1)$ symmetry can serve as a generator of texture zeros in
the Yukawa matrices, the maximal parameter reduction it can enforce
consistently with experimental data, and which type of textures can
arise. Throughout our analysis we assume two Higgs doublets carrying
different $U(1)$ charges, the minimum number required
to enforce symmetries of this type. 
Although generating quark Yukawa textures by imposing a $U(1)$
symmetry may seem an obvious pathway to explore, we are only aware
of a couple of studies in this direction
\cite{Ferreira:2010ir,Serodio:2013gka}.  These focus on the
classification of all combinatorially allowed textures compatible
with some (typically discrete) Abelian symmetry, with emphasis on
combinatorics rather than minimality.
% \footnote{Maximally restrictive textures in the lepton 
%   sector have been studied in \cite{Felipe:2016sya}.} 
Some of our findings are unexpected
and non-trivial: the maximum parameter reduction is to nine
non-vanishing entries which can only be arranged in $M_4 \oplus M_5$
structures, where the subscripts indicate the number of non-vanishing
entries in the pair of Yukawa matrices.  Among all the inequivalent
structures of this type, only two are consistent with a $U(1)$
symmetry.
Folding in the different possibilities for assigning flavour labels to
the matrix rows and columns gives rise to twelve different
`models'. Interestingly, in all cases the resulting $U(1)$ symmetry
has a QCD anomaly, and can thus be interpreted as a Peccei-Quinn (PQ)
symmetry.  Indeed, the idea that a PQ symmetry might have a
non-trivial relationship with flavour was proposed long ago by
Wilczek~\cite{Wilczek:1982rv} (see also
\cite{Geng:1988nc,Berezhiani:1989fp,Hindmarsh:1997ac}), and the
possible connection between axion models and the SM flavour puzzle
has recently triggered a revived interest~\cite{Bjorkeroth:2018dzu,%
  Reig:2018ocz,Linster:2018avp,Ahn:2018cau,Arias-Aragon:2017eww,%
  Ema:2016ops,Calibbi:2016hwq,Alanne:2018fns,Bertolini:2014aia,Celis:2014jua,%
  Ahn:2014gva,Cheung:2010hk,Albrecht:2010xh,%
  Appelquist:2006xd}.  Interestingly, in some cases only the light
quark generation contributes to the $U(1)$--QCD anomaly and, as
was recently shown in \cite{DiLuzio:2017ogq}, this feature opens up
the possibility that axion couplings to nucleons are sizeably weaker 
than what is generally assumed.

The paper is structured as follows.  In Section~\ref{sec:symmetries}
we derive a set of rules which provide the largest number of texture
zeroes, compatible with a generation-dependent $U(1)$ symmetry and
able to reproduce the observed quark masses and mixings.  In
Section~\ref{sec:predictions} we discuss the main phenomenological
consequences of these constructions, focusing in particular on the
predictions for axion couplings.  We conclude in
Section~\ref{sec:conclusions}. Ancillary results and numerical
fits are deferred to a set of appendices.

\section{\texorpdfstring{$U(1)$}{U(1)} symmetry and Yukawa textures}
\label{sec:symmetries} 

In order to accommodate an extra global $U(1)$ symmetry in the
renormalizable quark Yukawa Lagrangian, we assume the minimal content
of two Higgs doublets $H_{1,2}$, taking their hypercharges to be
$Y(H_{1,2}) = -1/2$, and which acquire vevs
$\langle H_{1,2}\rangle = v_{1,2}$ such that
$v_1^2+v_2^2\equiv v^2\simeq (174\,\mathrm{GeV})^2$.  We assume that
$H_{1,2}$ and the quark fields carry some new $U(1)$ global charge
$\mX$. 
As discussed at the end of this section, the method for
identifying viable textures does not rely on particular values of
the Higgs charges, which could be left generic.  On the other hand,
a convenient choice of placeholder charges can simplify the analysis
considerably. Hence we assign to the Higgs fields
respectively the charges $ \mX_{1,2} \equiv \mX(H_{1,2}) = \pm 1$,
which is a particularly convenient choice. We take the quark charges
to be generation-dependent, so that $H_{1,2}$ couple in a
generation-dependent way to the quark bilinears $\Qbar_i u_j$ and
$\Qbar_i d_j$ ($i,j=1,2,3$) where $Q_i$ denote the left-handed (LH)
quark doublets and $u_j,\,d_j$ the right-handed (RH) $SU(2)_L$
singlets. Here the labels $ i $, $ j $ do not refer to any particular
ordering, and $u$, $d$ should also be understood as dummy flavour
labels, so that permutations of the $i,j$ indices and
$u \leftrightarrow d$ relabeling can be performed freely to comply
with a consistent physical interpretation.  As will be explained at
the end of this section, this choice is without loss of generality as
far as the search for viable Yukawa structures is concerned.

To proceed, let us assign to the quark fields generic $U(1)$ charges
$\mX(\Qbar) = \{x,y,0\}$, $\mX(u) = \{a,b,c\}$, and
$\mX(d) = \{m,n,p\}$, where the charge of one quark field (here $Q_3$)
can be always set to zero by a redefinition of all charges,
proportional to baryon number $ B $.  The $U(1)$ charge structure of
the Yukawa bilinears reads
\begin{equation}
\label{eq:MdMu}
  \mX_{\Qbar u} =\pmatr{
  a+x & b+x & c+ x \\
  a+y & b+y & c+ y \\
  a & b & c
  }, 
\qquad 
  \mX_{\Qbar d} =\pmatr{
  m+x & n+x & p+ x \\
  m+y & n+y & p+ y \\
  m & n & p
  }.
\end{equation}
Here and in the corresponding Yukawa matrices, the LH doublets $Q_i$ label the rows, and the RH singlets
$u_i,d_i$ label the columns. It is straightforward to see that
consistency of $U(1)$ charge assignments yields constraints for the charge
difference between pair of entries in $\mX_{\Qbar u}$, $\mX_{\Qbar d}$,
\begin{equation}
\begin{aligned}
	\big(\mX_{\Qbar d}\big)_{ij} -
	\big(\mX_{\Qbar d}\big)_{ik} &= \Delta^d_{jk}\,, \\ 
	\big(\mX_{\Qbar u}\big)_{ij} -
	\big(\mX_{\Qbar u}\big)_{ik} &= \Delta^u_{jk}\,, \\
	\big(\mX_{\Qbar d}\big)_{ji} -
	\big(\mX_{\Qbar d}\big)_{ki} = 
	\big(\mX_{\Qbar u}\big)_{ji} -
	\big(\mX_{\Qbar u}\big)_{ki} &= \Delta^Q_{jk}\,,
\end{aligned}
\label{eq:Delta}
\end{equation}
that is, the differences in \eqn{eq:Delta} are independent of the
index $i$ and, for example, $ \Delta^Q_{12}=x-y$, $ \Delta^Q_{23}=y$,
etc. Clearly, the only non-zero entries in the Yukawa matrices
$M_{d,u}$ will be the ones for which
$\mX_{\Qbar d},\mX_{\Qbar u}=\pm 1$ so that a $U(1)$-invariant
coupling with one of the two Higgs is possible.  

We want to establish what is the maximum reduction in the number of
non-zero entries in $M_{d,u}$ that can be enforced by a $U(1)$
symmetry, since the corresponding reduction in the number of free
Yukawa parameters would yield models with enhanced predictivity.
Viable constructions must have
$U(1)$ charge assignments consistent with \eqn{eq:Delta}, as well as
with a set of phenomenological constraints: no massless quarks, no
vanishing mixing angles in the CKM matrix $V_\mathrm{CKM}$, and CP
violation from a complex phase in $V_\mathrm{CKM}$.  These conditions
can be formulated more precisely in terms of generic Yukawa matrices
with complex entries $M_d$, $M_u$.%
\footnote{
  For convenience, $M_{d,u}$ denote two Yukawa matrices multiplied by the dimensional
  parameter $v$ and with their non-zero entries appropriately
  rescaled by weight factors of $\sin\beta$ or $\cos\beta$, with
  $\tan\beta = v_2/v_1$. However, it is understood that $M_{d,u}$, and
  in particular their structures, are properly defined only
  at a scale well above the electroweak breaking scale.}  
We require:
\begin{itemize}
\item A non-vanishing commutator
  $\mathcal{D}\equiv \det[M_d M_d^\dagger,M_u M_u^\dagger] $.
  $ \mathcal{D} \neq 0 $ is a necessary condition for a non-vanishing
  Jarskog invariant
  $J \propto \mathrm{Im}[\mathcal{D}] $~\cite{Jarlskog:1985ht}, which
  in turn ensures that all mixing angles and the CP-violating phase in
  $V_\mathrm{CKM}$ are non-vanishing. Since with nine quark fields
  there are eight relative phase redefinitions that can be used to
  remove complex phases in the Yukawa parameters, 
  we can immediately conclude that a minimum of nine non-zero entries (of which eight can
  be made real) is a necessary condition for CP violation.%
  \footnote{The total parameter freedom can be further reduced by assuming symmetric
    matrices \cite{Giudice:1992an,Dimopoulos:1991za} in which case the
    number of free parameters is reduced while the number of
    non-vanishing entries is not. However, in the absence of a
    mechanism enforcing this condition 
    (e.g.~$\mathcal{C}$-parity in left-right symmetric models), 
    such an assumption is not justified.
  }
\item Non-zero determinants $\det[M_d] \neq 0$ and $\det[M_u] \neq 0$.
  This ensures no massless quarks, and has the immediate
  consequence that some structures are not viable. Denoting with a
  subscript the number of non-zero elements in a mass matrix, these
  are $M_0 \oplus M_9$, $M_1 \oplus M_8$ and $M_2 \oplus M_7$.
\item Consistency of the $U(1)$ charge assignment with \eqn{eq:MdMu}.
  In particular this implies that $M_3 \oplus M_6$ structures must be
  also discarded because, as shown in Appendix \ref{app:3p6}, they
  cannot be enforced by consistent $U(1)$ assignments.
\end{itemize} 

We conclude that the only viable structures with the minimum number of
nine parameters have the form $M_4 \oplus M_5$.  
In our study we collect the large number of possible
$M_4 \oplus M_5$ structures in equivalence classes containing pairs of
matrices that, for a fixed set of numerical inputs, yield the same mass
eigenvalues and CKM mixings. For example, independent permutations of
the columns in $M_4$ and $ M_5 $ affect only RH mixing, which are not SM
observables, while permuting the rows in both
matrices in the same way amounts to relabeling the quark doublets $Q_{1,2,3}$ with no effect
on $V_\mathrm{CKM}$.

Let us now proceed to identify the possible inequivalent Yukawa
textures.  Let us start with $M_4$, to which we assign the dummy
variable `$d$' so that the matrix of charges of the quark bilinears is
$\mX_{\Qbar d}$.  To ensure $\det[M_4]\neq 0$, the first three entries
can be arranged in $\frac{1}{3!} (9\cdot 4\cdot 1)$ ways.  That is,
the first entry can go anywhere (9 possibilities), the second in the
$2\times 2$ submatrix that does not contain the first entry, with the
position of the last entry then fixed.  The fourth entry can now go in
any of the six remaining empty positions (the combinatorial factor
does not change because this entry cannot contribute to the
determinant). We can then permute the columns in six ways and the rows
in six ways to get equivalent configurations, such that the final
number of inequivalent textures for $M_4$ is
  $\frac{1}{3!} (9\cdot 4\cdot 1) \cdot 6 \cdot \frac{1}{6\cdot 6} =
  1$.

Let us choose as the representative structure for $M_4$ three nonzero
entries on the diagonal plus one in $(M_4)_{12}$ to which we assign,
without loss of generality, a complex phase, i.e.
$(M_4)_{12} =|(M_4)_{12}|e^{i\alpha}$. 
$M_5$ will then be taken to be
a real matrix.  There are two possible charge assignments that can
realize this texture, depending on if we choose
$\diag(\mX_{\Qbar d}) = (s,s,-s)$ or
$\diag(\mX_{\Qbar d}) = (s,s,s)$, where $s=\pm 1$.
The remaining entries $\mX_{\Qbar d}$ can be filled up by requiring 
charge consistency and that no additional entries besides 
the diagonal and $(M_4)_{12}$ are allowed in $M_4$. 
This yields 
\begin{equation}
\begin{aligned}
	\mX_{\Qbar d} &= \begin{pmatrix}
		\bs & -\bs & -(k+2)s \\
		3s & \bs  & -ks \\
		(k+2)s & ks  & -\bs 
	\end{pmatrix}_{k\neq\pm 1,-3},
	\\
	\widetilde{\mX}_{\Qbar d} &= \begin{pmatrix}
		\bs & -\bs & -(k+2)s \\
		3s & \bs  & - ks \\
		(k+4)s & (k+2)s  & \bs 
	\end{pmatrix}_{k\neq\pm1,-3,-5}, 
\end{aligned}
\label{eq:M4}
\end{equation}
where for clarity a boldface $\bs$ has been used whenever the bilinears match a Higgs charge.  
Note that the opposite sign entry $(-s)$ cannot appear in one of the first two positions along the
diagonal, as this implies that if $(M_4)_{12}\neq 0$ then also
$(M_4)_{21}$ is non-vanishing.

Let us now study $M_5$. Again, without loss of generality we fill the
diagonal with three nonzero entries. The remaining two entries can be
assigned according to three possible configurations: $(i)$ block
diagonal, in which one entry has no other non-vanishing entries in the
row and column to which it belongs, $(ii)$ a row or a column filled
with three entries, $(iii)$ `democratic'  textures that do not belong
to $(i)$ or $(ii)$; see the following examples,
\begin{equation}
	(i)\!:~ 
	M_5 = \begin{pmatrix}
		\times &   &  \\
	    & \times & \times \\
	    & \times & \times 
	\end{pmatrix},
	\qquad
	(ii)\!:~ 
	M_5 = \begin{pmatrix}
		\times &   & \times \\
	    & \times & \times \\
	    &   & \times  
	\end{pmatrix},
	\qquad 
	(iii)\!:~ 
	M_5 = \begin{pmatrix}
		\times &  \times & \\
	    & \times & \times \\
	    &   & \times 
	\end{pmatrix}.
\label{eq:M5i}
\end{equation}
\sloppy Block diagonal textures $(i)$ are not viable because, as we will now
argue, they yield $\det[M_4M_4^\dagger,M_5M_5^\dagger] =0$.  The
structure of $M_4$ preserves a flavour symmetry for $Q_3$, see
\eqn{eq:M4}.  In order to break this symmetry and mix $Q_3$ with the
other quark doublets then the $2\times 2$ block must involve the $Q_3$
row, as in case~(\ref{eq:M5i}$\,i$).  Let us now write
$M_4M_4^\dagger = \Phi S_4 \Phi^*$ with
$\Phi = \diag(e^{i \alpha},1,1)$ and $S_4$ a real symmetric
matrix. $M_5M_5^\dagger = S_5$ is also real symmetric and satisfies
$\Phi S_5 \Phi^* = S_5$. It follows that we can write the commutator
as
\begin{equation}
	[M_4M_4^\dagger,M_5M_5^\dagger] = \Phi [S_4,S_5]\Phi^*\,.  
\end{equation}
The commutator of two symmetric matrices is an antisymmetric
matrix, and the determinant of an antisymmetric matrix of odd
dimension vanishes. The other block-diagonal texture, with
off-diagonal entries assigned to $(M_5)_{13}$ and $(M_5)_{31}$, can be
brought to the same form as (\ref{eq:M5i}$\,i$) 
by permuting the first two columns and rows in both $M_4$ and $M_5$, and then the same
argument applies.  

Under the requirement of only five non-vanishing entries, filling a
row or a column with three entries, case~$(ii)$, clashes with $U(1)$
charge consistency.  This can be understood as follows: as $s=\pm 1$ are 
the only possible choices for the charges of non-vanishing
entries, at least two of the three allowed entries in the row/column
must have the same charge, take $\mX_{13}=\mX_{33}$ in example
(\ref{eq:M5i}\,$ii$). The difference between their charges then
vanishes.  By taking a second row/column with a non-vanishing entry in
the same column/row position ($\mX_{11}$ for the case at hand) we see
that to match the vanishing of the corresponding charge difference
there must be another entry with the same charge (here $\mX_{31}$) which implies a sixth non-vanishing entry.

The remaining possibilities are thus the democratic textures of type $(iii)$ (and permutations), 
for which the only consistent possibility for charge assignments is
\begin{equation}
	\mX_{\Qbar u} = \begin{pmatrix}
		\mathbf{1} &  \mathbf{-1}  &   -3 \\
		3 & ~~ \mathbf{1} &  \mathbf{-1}  \\
		5 & ~~ 3 & ~~ \mathbf{1}
	\end{pmatrix}.
\label{eq:XQu}
\end{equation}
Here we have used explicit numerical values for the bilinear charges
since the possibility of a relative minus sign with respect to the
charges in $\mX_{\Qbar d}$ is already accounted for in \eqn{eq:M4} by
$s=\pm1$. \Eqn{eq:XQu} gives the row charge differences
$\Delta^Q_{12}=\Delta^Q_{23}=-2$ which, for consistency, should be
respected also by some permutation of the matrices in \eqn{eq:M4}.
For $\mX_{\Qbar d}$ we obtain $s=+1$ and $k=3$ by straightforward inspection.
However, there is also another possibility which is obtained by
permuting the first and third row and gives $s=-1$ and $k=3$.  All other
permutations yield either $s\neq \pm1$ or $k=\pm 1,-3$ and hence can
be discarded. 
For $\widetilde{\mX}_{\Qbar d}$ we obtain $s=+1$ and $k=1$ which
is forbidden, and similarly none of the additional five
possibilities obtained by permuting the rows is viable. We conclude
that there are only two possible charge assignments compatible 
with the requirement of maximal reduction in the number of Yukawa
parameters (nine non-vanishing Yukawa couplings) and with $U(1)$
charge consistency.
These are \eqn{eq:XQu} (which gives $M_5$) together with 
\begin{equation}
	\mX_{\Qbar d}^{(1)} = \begin{pmatrix}
		\mathbf{1} & \mathbf{-1} & -5 \\
		3 & ~~ \mathbf{1} & -3  \\
		5 & ~~ 3 & \mathbf{-1}
	\end{pmatrix}, 
	\qquad
	\mX_{\Qbar d}^{(2)} = \begin{pmatrix}
		-5 & -3 & ~~ \mathbf{1} \\
		-3 & \mathbf{-1} 	& ~~ 3 \\
		\mathbf{-1}& ~~ \mathbf{1} & ~~ 5
	\end{pmatrix},
\label{eq:XQd}
\end{equation}
which give $M_4^{(1,2)}$.
We denote the combined Yukawa structures by $\mathcal{T}_{1,2} = M_4^{(1,2)}
\oplus M_5$, corresponding to the structures
\begin{equation}
    \mathcal{T}_1 = 
    \pmatr{\times & \times & 0 \\ 0 & \times & 0 \\ 0 & 0 & \times} \oplus 
    \pmatr{\times & \times & 0 \\ 0 & \times & \times \\ 0 & 0 & \times},
    \qquad
    \mathcal{T}_2 = 
    \pmatr{0 & 0 & \times \\ 0 & \times & 0 \\ \times & \times & 0} \oplus 
    \pmatr{\times & \times & 0 \\ 0 & \times & \times \\ 0 & 0 & \times} ,
\label{eq:textures}
\end{equation}
where the ``$\times$'' denote the non-vanishing Yukawa entries.

As was mentioned at the beginning of this section, the choice
$\mX_{1,2}=\pm 1$ for the Higgs charges represents just a
simplification to carry out the analysis (since
$\mX(H_2^\dagger)=\mX(H_1)$), but it does not imply any restriction
for the correct identification of the viable textures, and the same
result would have been obtained with generic Higgs charges. To see
this, let us choose the charge normalization
$\mX(H_1^\dagger H_2) = \mX_2-\mX_1 = 1$, so that
$\mX_1=-s^2_\beta \equiv -\sin^2\beta$ and
$\mX_2=c^2_\beta \equiv \cos^2\beta$.  \Eqns{eq:XQu}{eq:XQd} can be
rewritten in terms of these generic charges by making the following
substitutions: in \eqn{eq:XQu}, $ \mathbf{1} \to -\mX_1 $ and
$ \mathbf{-1} \to - \mX_2 $, while in \eqn{eq:XQd}, to match the
conjugate Higgs doublets, $ \mathbf{-1} \to \mX_1$ and
$ \mathbf{1} \to \mX_2 $.%
\footnote{ This fixes the flavour label $ u $ to refer unambiguously
  to the up sector; if $ \mX_{\Qbar u} $ is associated instead with
  the down quark bilinears, we have the substitutions: in
  \eqn{eq:XQu}, $ \mathbf{1} \to \mX_2 $ and
  $ \mathbf{-1} \to \mX_1 $, while in \eqn{eq:XQd},
  $ \mathbf{-1} \to -\mX_2 $ and $ \mathbf{1} \to -\mX_1 $.  }
Comparing these entries with the corresponding entries in
\eqn{eq:MdMu} and solving the linear system of nine equations
determines the values of the individual charges $\{x,y\}$,
$\{a,b,c\}$, and $\{m,n,p\}$, which in turn allows us to reconstruct
the complete charge matrices in terms of $s^2_\beta,\,c^2_\beta$.

\section{Phenomenological constraints and predictions}
\label{sec:predictions}

As we have argued, the maximal reduction in the number of Yukawa
parameters yields nine real parameters and one complex phase, thus
matching one-to-one the number of independent observables: six quark
masses, three mixing angles, and one CKM phase. This ensures that any
$M_4\oplus M_5$ Yukawa texture can successfully fit the experimental
data and, most importantly, that any other quantity which
depends on the Yukawa matrices is predicted. More precisely, while
in the SM only one combination of the LH and RH quark mixing matrices
$\mathcal{V}_{L,R}^{u,d}$ is observable, namely
$V_\mathrm{CKM}=\mathcal{V}_L^u{\mathcal{V}_L^d}^\dagger$, new physics (NP) 
processes might be sensitive to other combinations of the 
diagonalizing matrices. In the case at hand, the spontaneous breaking 
of the $U(1)$ symmetry at the scale $f_a$ will lead to a QCD axion, whose 
couplings to quarks (in particular the flavour-violating ones) 
depend on the individual mixing matrices and hence can be univocally predicted, 
modulo an overall factor $1/f_a$ suppressing the coupling strength.

\subsection{Reconstruction of the Yukawa matrices}
\label{sec:fit}

The exact match between the numbers of fundamental Yukawa parameters
and flavour observables ensures that a complete reconstruction of
the fundamental Yukawa matrices in terms of measured quantities is always possible.
This is an important step in computing the individual
$\mathcal{V}_{L,R}^{u,d}$ mixing matrices that control NP processes.
However, in practice, carrying out such a task is not completely
straightforward, and we will now illustrate the main steps that allow
for $M_{u,d}$ reconstruction.

Clearly, identifying $M_4$ with the down or with the up quark Yukawa
matrix will yield different physics. We also note that, for example, 
$M_4^{(1)}$ has a single entry in the third row and third column, which can be
arbitrarily identified with one of the six quark flavours $d,s,b$, or
$u,c,t$.  
Since this particular flavour will not mix with the other two quark 
flavours of the same electric charge, different choices will yield 
qualitatively different physics. 
We will refer to this flavour as the ``sequestered'' quark.
Meanwhile, a different labeling for the remaining entries, for which 
all the quarks of the same electric charge do mix, is equivalent to 
a trivial reshuffling of the corresponding quark
labels. In summary, we have two different structures, 
$\mathcal{T}_{1,2}$ in \eqn{eq:textures}, two ways to assign the up 
and down quarks to $M_{4,5}$, and three ways to identify the 
sequestered quark, for a total of 12 different models.  For each one 
of the two textures $\mathcal{T}_{1,2} $ we label the six possible 
models by a superscript $(q)$, where $q = u$, $c$, $t$, $d$, $s$, $b$ 
labels the sequestered quark in $M_4$, e.g. $ \mathcal{T}_{1}^{(t)} $ 
labels the model where the $(3, 3)$ element of $ M_4 $ corresponds to the top quark.

As above, we use the labels $ M_d = M_4 $ and $ M_u = M_5 $ with the
understanding that $u$ and $d$ will remain dummy labels until the
analytical expressions are matched to the flavour observables.  Let us
study the case of $\mathcal{T}_1 = M_4^{(1)}\oplus M_5$ (the same analysis 
can be straightforwardly carried out for $\mathcal{T}_2$).  
The entries of the corresponding Yukawa matrices are labeled as
\begin{equation}
	M_d = \pmatr{
		m^d_{11} & m^d_{12} e^{i \alpha} & 0 \\
		0 & m^d_{22} & 0 \\
		0 & 0 & m^d_{33}
	}, \qquad
	M_u = \pmatr{
		m^u_{11} & m^u_{12}  & 0 \\
		0 & m^u_{22} & m^u_{23}  \\
		0 & 0 & m^u_{33} 
	},
\label{eq:MduFit}
\end{equation}
where $m_{ij}^{u,d}$ are real parameters.  Let us define the Hermitian
matrices $\mathcal{D}= M_d M_d^\dagger$ and
$\mathcal{U}= M_u M_u^\dagger$.  They are diagonalized by the two unitary
matrices $\uV{L}{u,d}$, such that
\begin{equation}
  \mathcal{D} = \uV{L}{d\dagger}\,\hat{M}_d^2\, \uV{L}{d}, \qquad
  \mathcal{U} = \uV{L}{u\dagger}\,\hat{M}^2_u\,  \uV{L}{u},
\label{eq:Md2Mu2}
\end{equation}
where $ \hat{M}_{d} = \diag(m_{d1},m_{d2},m_{d3}) $ and $ \hat{M}_{u} =
\diag(m_{u1},m_{u2},m_{u3}) $  are diagonal Yukawa matrices proportional to 
the physical masses.  
$ \uV{L}{d} $ takes the form
\begin{equation}
	\uV{L}{d} = \pmatr{
	c_\tau & - e^{i \alpha} s_\tau & 0 \\
	e^{-i \alpha} s_\tau  & c_\tau & 0 \\
	0 & 0 & 1
	}.
\end{equation}
where $c_\tau = \cos\tau$, $s_\tau = \sin\tau$, and the phase has been fixed by means of the first
relation in \eqn{eq:Md2Mu2}.  The CKM matrix is defined as $ V_\mathrm{CKM} \equiv \uV{L}{u} \uV{L}{d\dagger} $, and thus  we can write $ \mathcal{U} = \uV{L}{d\dagger} V_\mathrm{CKM}^\dagger \hat{M}_u^2 V_\mathrm{CKM} \uV{L}{d} $.
The texture zeroes in $ M_u $ imply $\mathcal{U}_{13} = \mathcal{U}_{31} = 0 $. Explicitly,
\begin{equation}
  \mathcal{U}_{13} = 
  c_\tau  \left(V_\mathrm{CKM}^\dagger {\hat M}_u^2V_\mathrm{CKM}\right)_{13} 
  + e^{i \alpha} s_\tau \left(V_\mathrm{CKM}^\dagger {\hat M}_u^2V_\mathrm{CKM}\right)_{23}
  = 0, 
\end{equation}
which yields 
\begin{equation}
	e^{i \alpha} \tan \tau = - \frac{
          \left(V_\mathrm{CKM}^\dagger {\hat
              M}_u^2V_\mathrm{CKM}\right)_{13}}{
  \left(V_\mathrm{CKM}^\dagger {\hat M}_u^2V_\mathrm{CKM}\right)_{23}}\,.
\label{eq:umixing}
\end{equation}
This completely determines $\uV{L}{d}$ in terms of measured quantities.
With $ \uV{L}{d} $ fixed it is straightforward to determine the couplings $ m^d_{ij} $ by means of the first relation in \eqn{eq:Md2Mu2}. We obtain 
\begin{equation}
% \medmuskip=0mu
% \thinmuskip=0mu
\thickmuskip=0mu
	m^d_{33}= m_{d3} , ~~ 
	m^d_{22}= \sqrt{ m_{d2}^2 c_\tau^2 +  m_{d1}^2 s_\tau^2}, ~~ 
	m^d_{12}= \frac{(m_{d2}^2 -m_{d1}^2)c_\tau s_\tau}{m^d_{22}}, ~~
	m^d_{11}= \sqrt{m_{d1}^2 c_\tau^2 +  m_{d2}^2 s_\tau^2 - (m^d_{12})^2}\,. 
\end{equation}
$\mathcal{U}$ is also completely determined in terms of known
parameters, as $ \mathcal{U} = \uV{L}{d\dagger} V_\mathrm{CKM}^\dagger \hat{M}_u^2
V_\mathrm{CKM} \uV{L}{d} $.
By comparing this expression with $ \mathcal{U} = M_u M_u^\dagger$ (with $M_u$ as in \eqn{eq:MduFit}),
the fundamental couplings $ m^u_{ij} $ can be easily determined, at least numerically, according to
\begin{equation}
m^u_{33} = \sqrt{\mathcal{U}_{33}} ,  \quad
m^u_{23} = \frac{\mathcal{U}_{23}}{m^u_{33}} ,\quad 
m^u_{22} = \sqrt{\mathcal{U}_{22} - (m^u_{23})^2} , \quad 
m^u_{12} = \frac{|\mathcal{U}_{12}|}{m^u_{22}} , \quad
m^u_{11} = \sqrt{\mathcal{U}_{11} - (m^u_{12})^2} .
\end{equation}
Note that in general $\mathcal{U}_{12}$, written in terms of
$\uV{L}{d}$ and $V_\mathrm{CKM}$, will not turn out to be real; a
real value, in agreement with the matrix in \eqn{eq:MduFit}, can be
obtained by shifting the phase in $ M_d $ as
$ \alpha\to\alpha-\arg\mathcal{U}_{12} $. 
Finally, the RH mixing matrices $ \uV{R}{d,u}$ can be straightforwardly obtained
(numerically) by diagonalization of the Hermitian matrices
$M_d^\dagger M_d $ and $M_u^\dagger M_u $.

\subsection{Observable parameters and stability}
\label{sec:input} 

Since the Higgs doublets carry $U(1)$ charges it is clear that the
symmetry will be spontaneously broken. In fact, as we will see below,
the breaking must occur at a scale much larger than the electroweak
scale (via a SM-singlet scalar field $\phi$) to sufficiently suppress
the couplings of the $U(1)$ Goldstone boson to the SM fields. After
$U(1)$ breaking the texture zeroes are no longer protected, and will
be lifted to non-zero values by renormalization group (RG) running
effects.%
\footnote{
  In principle the texture zeroes can also be lifted by
  higher-dimensional operators that preserve the $U(1)$ symmetry, or
  even that violate it if they are gravity-induced
  \cite{Holman:1992us,Kamionkowski:1992mf,Barr:1992qq}. 
  Assuming a cutoff scale for these operators of
  $ \mathcal{O}(M_\mathrm{Planck}) $ renders their effect subdominant
  with respect to the effects of RG running.
}
This means that, for
consistency, the non-zero entries in the Yukawa matrices should be
determined in terms of the high-scale values of the SM observables. RG
running of the SM parameters from low energy to various high-energy
scales has been performed by various
groups~\cite{Fusaoka:1998vc,Xing:2007fb,Xing:2011aa,Antusch:2013jca}.
While the exact scale of $ U(1) $ breaking is to some extent
arbitrary, we anticipate that it has to be compatible with axion
phenomenology, thus fixing the value of the symmetry-breaking order
parameter $ v_\phi \sim f_a \gtrsim 10^{8}$ GeV.  We then use the
values for the quark masses at $ \mu = 4 \times 10^{12} $ GeV given in
\cite{Xing:2011aa}. 
In this reference, the corresponding high scale values of the CKM mixing
angles and CP phase are not given. However, it is known that these
quantities do not run much. We have hence adopted the results given in
\cite{Antusch:2013jca} which correspond to a scale $ \mu = M_Z $.  The
values of our input parameters are given in Table \ref{tab:exp}.

\begin{table}[ht]
\centering
\renewcommand{\arraystretch}{1.2}
\begin{tabular}{c D{@}{\,}{9,9} c@{\hskip3pt} c D{@}{\,}{9,9}}
\toprule
	Observable & \multicolumn{1}{c}{Value} && Observable & \multicolumn{1}{c}{Value} \\
\cmidrule{1-2} \cmidrule{4-5}
	$ m_u \,/$MeV &  0.61 	@ ^{+0.19}_{-0.18} &
	& $ \theta_{12} $ & 0.22735 @ \pm 0.00072 \\
	$ m_c \,/$GeV &  0.281 	@ ^{+0.02}_{-0.04} &
	& $ \theta_{13} $ & 0.00364 @ \pm 0.00013 \\
	$ m_t \,/$GeV &  82.6 	@ \pm 1.4 &
	& $ \theta_{23} $ & 0.04208 @ \pm 0.00064 \\
	$ m_d \,/$MeV &  1.27 	@ \pm 0.22 &
	& $ \delta $ & 1.208 @ \pm 0.054  \\
	$ m_s \,/$MeV &  26 	@ ^{+8}_{-5} &&& \\
	$ m_b \,/$GeV &  1.16 	@ ^{+0.07}_{-0.02} &&& \\
\bottomrule
\end{tabular}
\caption{Input values of the quark masses and CKM parameters 
taken from \cite{Xing:2011aa,Antusch:2013jca} (see text).}
\label{tab:exp}
\end{table}

As noted above, for each of the two textures identified in Section
\ref{sec:symmetries} there are six physically distinct arrangements of
the quark flavours, for a total of twelve sets of input parameters,
all of which reproduce the SM data but in general yield different NP
effects.
The complete sets of numerical solutions are given in Appendix~\ref{app:fits}.
It is natural to ask to what extent these solutions
  are numerically fine-tuned, or in other words how stable they are
  under small perturbations, as for example when the reference values
  of the observables are varied within their respective experimental
  uncertainties. Equivalently, it would be desirable if Yukawa
  matrices that do not differ too much from the exact solutions 
  would still yield acceptable values for the observables.

The simplest mechanism that can yield large fluctuations is
  when a small number arises from a tuned cancellation between the
  values of two large parameters. Clearly this requires at least one
  pair of parameters with sufficiently close values. Since there are
  no pairs of experimental observables with close values, nor among the 
  numerical entries in the matrices in
  Appendix~\ref{app:fits}, we conclude that there is no such simple 
  source of instability in the correspondence between observables 
  $ \mathcal{O} $ and fundamental parameters $ p $ of the Yukawa matrices.
  This is true for both the direct ($ \mathcal{O} = \mathcal{O}(p) $) and 
  inverse ($ p = p(\mathcal{O}) $) correspondence.
  However, given that the equations for the inverse problem are
  highly nonlinear, more complicated sources of fine-tuned
  cancellations are possible. A robust way to assess if
  fine-tuning is present in the correspondences is to define, 
  analogous to the Barbieri-Giudice measure of fine-tuning \cite{Barbieri:1987fn},
  the quantities
\begin{equation}
  Q_{ij} = \left| \frac{p_j}{\mathcal{O}_i} \frac{\Delta \mathcal{O}_i}{\Delta p_j} \right|,
\end{equation}
expressing the relative change in the value of an observable
$\mathcal{O}_i$ for a given relative change in the value of a
fundamental parameter $p_j$.  
If, for any $ i $, $ j $, 
$ Q_{ij} \gg 1 $ (when varying $ p_j $) or 
$ Q_{ij} \ll 1 $ (when varying $ \mathcal{O}_i $),
we expect a tuned solution.
We have verified that for all the numerical solutions listed in Appendix~\ref{app:fits},
$1/2 \lesssim Q_{ij} \lesssim 2 $ for all $i,j=1,2,\dots,10$, confirming that 
the numerical solutions of the direct and inverse
problems are remarkably stable, and absent of fine-tunings.

It is well known that models with two Higgs doublets with
generic couplings to the fermions can give rise to scalar-mediated
flavour-changing neutral currents (FCNCs), for which stringent
experimental bounds exist. These effects are ubiquitous in
multi-Higgs doublets models and not specific to our scenario, hence
their phenomenology, and various mechanisms to keep them under
control, are well documented in the literature. For simplicity we
will assume here that the Higgs sector can realize the decoupling
limit~\cite{Gunion:2002zf,Bernon:2015qea}, wherein the lightest
Higgs mass eigenstate has SM-like couplings, while the other
scalars are much more heavy and sufficiently decoupled.%
\footnote{
Another popular approach to suppress scalar-mediated FCNCs is
Yukawa alignment, where for a given fermion sector the Yukawa
coupling matrices associated with each Higgs are simultaneously
diagonalizable.  Specific realizations using flavour symmetries
have been considered in
\cite{Pich:2009sp,Serodio:2011hg,Varzielas:2011jr,Celis:2014zaa}.
This strategy, however, cannot be straightforwardly implemented in
our case.}

\subsection{The \texorpdfstring{$U(1)$}{U(1)} flavour symmetries are Peccei-Quinn symmetries}
\label{sec:PQsymmetries}

With a field content consisting only of the SM and two
Higgs doublets, the $U(1)$ flavour symmetry gets spontaneously
broken by the vevs $v_{1,2}$.  To identify the physical $U(1)$
Goldstone mode we must ensure that it is not mixed with the $Z$ boson, 
or more precisely with the Goldstone mode of $U(1)_Y$ of hypercharge. 
This fixes the ratio of the two Higgs charges to
$\mX_1/\mX_2 = - \tan^2 \beta$ \cite{Dine:1981rt}, where
$ \tan \beta = v_2/v_1$ is a free parameter.
With the charge normalization $\mX_1 = -s^2_\beta$ and
$\mX_2 = c^2_\beta$ already introduced at the end of Section
\ref{sec:symmetries}, the $U(1)$ charges of the quark bilinears in
\eqns{eq:XQu}{eq:XQd} correspond to the following charge assignments:
% \begin{table}[ht]
\begin{equation}
\renewcommand{\arraystretch}{1.2}
\centering
\begin{tabular}{|c|c|c|c|}
\hline
Texture & $ \mX(Q) $ & $ \mX(u) $ & $ \mX(d) $ \\[0.4ex]
\hline
$ \mathcal{T}_1^{(d,s,b)} $ & 
\multirow{4}{*}{$ \{2,1,0\} $} &
\multirow{2}{*}{$ \{3-c^2_\beta,2-c^2_\beta,1-c^2_\beta\} $} &
$ \{2+c^2_\beta,1+c^2_\beta,-1+c^2_\beta\} $
\\[0.4ex]
\cline{1-1}\cline{4-4}
$ \mathcal{T}_2^{(d,s,b)} $ & 
 & 
 & 
$ \{-1+c^2_\beta,c^2_\beta,2+c^2_\beta\} $
\\[0.4ex]
\cline{1-1}\cline{3-4}
$ \mathcal{T}_1^{(u,c,t)} $ & 
 & 
$ \{3-c^2_\beta,2-c^2_\beta,-c^2_\beta\} $ & 
\multirow{2}{*}{$ \{2+c^2_\beta,1+c^2_\beta,c^2_\beta\} $}
\\[0.4ex] 
\cline{1-1}\cline{3-3}
$ \mathcal{T}_2^{(u,c,t)} $ & 
 & 
$ \{-c^2_\beta,1-c^2_\beta,3-c^2_\beta\} $ & 
\\[0.4ex]
\hline
\end{tabular}
% \end{table}
\label{eq:finalcharges}
\end{equation}

The QCD anomaly number is given by%
\begin{equation}
	2N = \sum_i \left[ \mX(u) + \mX(d) - 2\mX(Q) \right]_i ,
\label{eq:anomalynumber}
\end{equation}
so that we obtain $ 2N (\mathcal{T}_1) = 2(\mX_2-\mX_1)=2 $, and
$ 2N (\mathcal{T}_2) =(\mX_2-\mX_1)= 1 $. 
In both cases there is an anomaly; the two $U(1)$ flavour 
symmetries therefore have the correct properties for
being identified with PQ symmetries, and in turn the Goldstone mode of
the new global symmetry can be identified with an axion. 
As in the usual Dine-Fischler-Srednicki-Zhitnitsky
(DFSZ)~\cite{Dine:1981rt,Zhitnitsky:1980tq} model, this axion can be 
compatible with low energy phenomenology by rendering it \textit{invisible}, 
by introducing a new scalar field $\phi$,
a singlet under the SM gauge group but carrying a $U(1)$ charge
$\mX_\phi$, and acquiring a large vev $v_\phi \gtrsim 10^8$ GeV. In
this way $v_\phi$ becomes the dominant order parameter for the
spontaneous breaking of $U(1)$, the QCD axion $a(x)$ dominantly
emerges form the phase of the SM singlet
$\phi \supset \tfrac{v_\phi}{\sqrt{2}} e\, ^{i a / v_\phi}$, and all
the axion couplings get suppressed by $1/v_\phi$.

The periodicity of the anomaly under a $U(1)$ transformation of the
quark fields $q \to e^{i \mX_q \theta} q$ relative to the periodicity of
the axion potential under the shift $a \to a + v_\phi \mX_\phi \theta$
determines the number of inequivalent minima of the axion
potential, i.e. the number of domain walls
$N_\mathrm{DW} = 2N / \mX_\phi$ that form in the early Universe around
the QCD phase transition.  The same quantity also enters in the
definition of the axion decay constant $f_{a} = v_\phi/N_\mathrm{DW}$.
In turn, $\mX_\phi$ can take the two values, $\mX_\phi= 1$ or
$\mX_\phi= 1/2$, depending on whether the renormalizable non-Hermitian
operator that couples the scalar singlet to the Higgs doublets is of
dimension three $ (H_2^\dagger H_1 \phi)$ or four $(H_2^\dagger H_1 \phi^2)$. 
Thus we have the following possibilities:
$N_\mathrm{DW}(\mathcal{T}_1) = 2$ or $4$, and
$N_\mathrm{DW}(\mathcal{T}_2)= 1$ or $2$.

\subsection{Axion couplings}
\label{sec:axioncouplings}

The axion couplings to quarks are given by
\begin{equation}
	\mathcal{L}_{aq} =  
		\frac{\partial_\mu a}{2 f_a}
	    \Big[ \bar{\mathbf{u}} \gamma^\mu \cVec^u \mathbf{u} + \bar{\mathbf{u}} \gamma^\mu \gamma_5 (\cAx^u-F^u) \mathbf{u} 
		+ \bar{\mathbf{d}} \gamma^\mu \cVec^d \mathbf{d} + \bar{\mathbf{d}} \gamma^\mu \gamma_5 (\cAx^d-F^d) \mathbf{d} \Big] ,
\label{eq:Laq}
\end{equation}
where $\mathbf{u} = (u,c,t) $ and $\mathbf{d} = (d,s,b) $ are vectors
of the quark mass eigenstates.  $F^{u,d} = \diag(f_{u,d},0,0)$, with
$f_u + f_d = 1$, are model-independent contributions to the light
quark couplings, originating from the $aG\tilde G$ term.
Defining $z = m_u/m_d$ and choosing $f_u = 1/(1+z)$ avoids tree-level axion-pion mixing 
(see e.g.~\cite{diCortona:2015ldu}).
The coupling matrices $\cVec^q$ and $\cAx^q$, for $q = u, d$, are given by
\begin{equation}
  \cVec^q = \frac{\mX_\phi}{2 N} \left( \uV{R}{q} \mX_{q_R} \uV{R}{q\dagger} + \uV{L}{q} \mX_{Q_L} \uV{L}{q\dagger} \right) , \qquad
  \cAx^q = \frac{\mX_\phi}{2 N} \left( \uV{R}{q} \mX_{q_R} \uV{R}{q\dagger} - \uV{L}{q} \mX_{Q_L} \uV{L}{q\dagger} \right) .
\label{eq:VfAf}
\end{equation}
Note that as $ V^q $ and $ A^q $ refer to couplings between mass
eigenstate quarks, the rows and columns are now explicitly ordered,
and we choose the usual ordering according to mass hierarchy,
e.g.~$ \cVec^u_{12} $ always refers to the axion coupling to an up and
a charm quark.  The axion couplings to quarks depend both on LH
($ \uV{L}{q} $) and RH ($ \uV{R}{q} $) mixing matrices.
Moreover, as the charges $ \mX_q $ are not universal, $ \cVec^q $ and $ \cAx^q $ are non-diagonal, leading to
flavour violation (FV).

The off-diagonal couplings can be probed in decays of heavy mesons; a
recent review of current experiments and their constraints on the PQ
scale is found in \cite{Bjorkeroth:2018dzu} (see in particular Table 2 of
that work).  In the current generation of experiments, searches for
kaon decays of the type $ K^+ \to \pi^+ a $ provide the best
sensitivity to a flavour-violating QCD axion.  The NA62 experiment at
the CERN SPS, whose primary goal is measuring
$ \mathrm{Br}(K^+ \to \pi^+ \nu \bar{\nu}) $, could potentially probe
scales up to $ f_a \gtrsim 10^{12} $ GeV.  Conversely, if we wish to
avoid the strongest bound from FV, we must ensure that, for either the
down or the strange quark, the FV interactions with the axion are 
particularly strongly suppressed.  From the discussion in
Section~\ref{sec:fit} it is clear that this can be obtained by
assigning the down quarks to $M_4^{(1,2)}$, and by choosing $ d $ or $ s $ as
the sequestered quark, which will then mix with the other same-type
quarks only due to subleading RG effects, which eventually lift the zero texture.
If, for example, the sequestered quark is the strange, both the
$ \bar{d}sa $ and $\bar{s}ba $ couplings would be
strongly suppressed, which in turn a provides a strong suppression 
of $K \to \pi a$ and $ B \to K a $ decays.  
$ B \to \pi a $ decays are still allowed at the leading order, yielding a 
limit $ f_a \gtrsim 10^8 |\cVec^d_{31}|$ GeV, which is safely
compatible the traditional axion window.

\begin{figure}[ht]
\centering
	\includegraphics[scale=0.8]{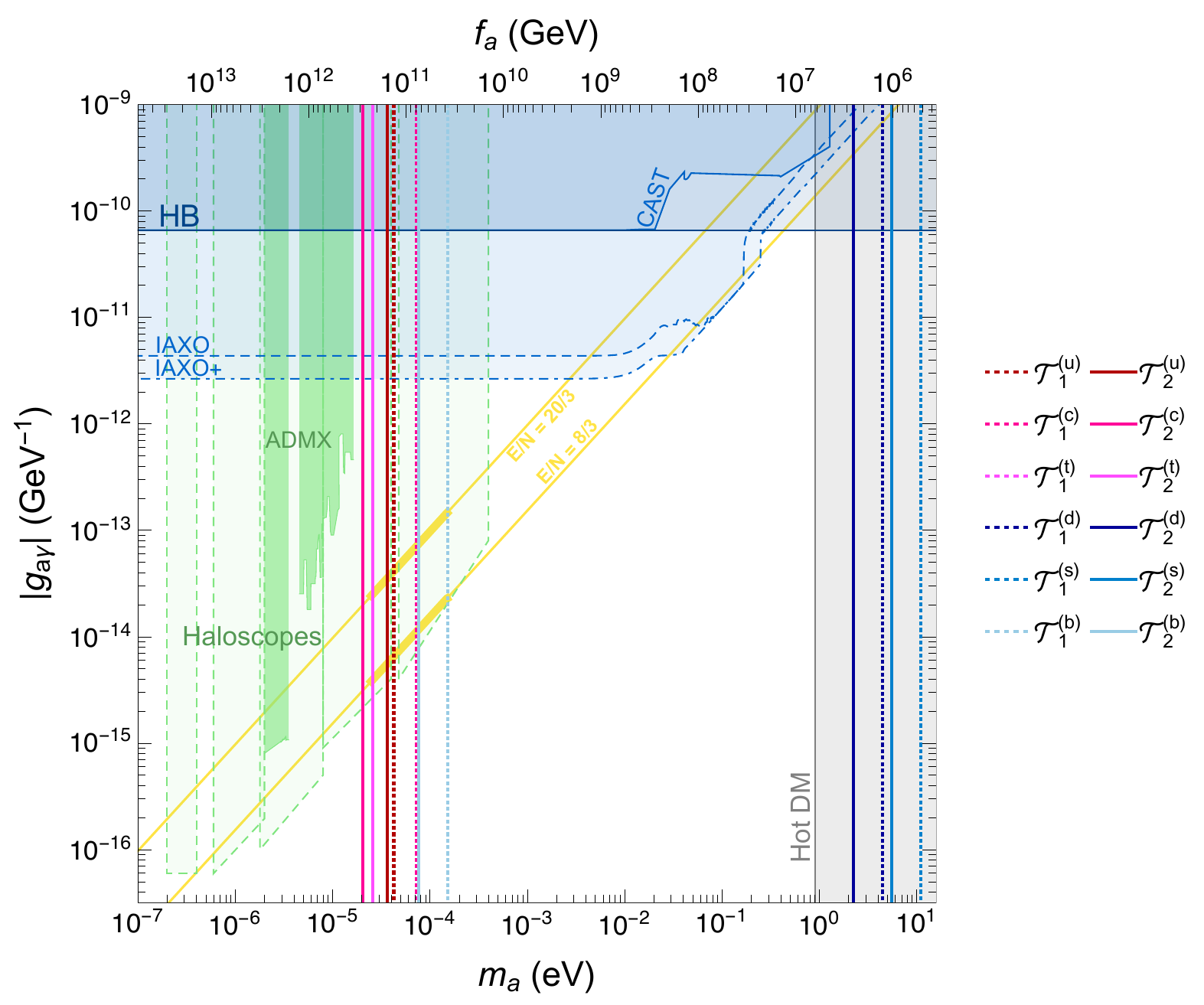}
        \caption{%
            Bounds on the axion-photon coupling $ g_{a \gamma} $ and
            axion mass $ m_a $.  The vertical colored lines denote
            the upper bound on $ m_a $ from FV, for the two different
            textures and quark assignments.  Also shown are
            astrophysical/cosmological bounds, and experimental
            sensitivities from helioscopes (in blue) and haloscopes
            (in green); see the text for more details. 
            The prediction for $ g_{a\gamma} $
            for the models discussed in the text, in which the PQ
            symmetry acts also on the lepton sector, lie between the two 
            oblique yellow lines, where the highlighted segment indicates 
            the favored mass region for axion DM.}
\label{fig:bounds}
\end{figure}

Constraints on the axion mass and axion-photon coupling
$ g_{a\gamma} $ are plotted in Fig.~\ref{fig:bounds}.  Experimental
bounds from FV are denoted by vertical lines, each referring to one
texture, with couplings fixed by flavour data (see Table
\ref{tab:fitandfvcouplings} of Appendix \ref{app:fits}).  Of the
twelve lines plotted, eight constrain $ m_a \lesssim 10^{-4} $ eV,
corresponding to the experimental bound on $ K^+ \to \pi^+ a $.
The remaining four lines correspond to the cases when either the
  $s$ or the $d$ quarks are sequestered, and $K \to \pi a$ transitions
  are only induced by the RG effects on the Yukawa matrices.  For
  $\mathcal{T}^{(s)}_{1,2}$ the bounds from $ B \to K a $ or
  $ B \to \pi a $ decays are typically stronger than the ones
  from RG-induced $K$ decays, 
  while for $\mathcal{T}_{1,2}^{(d)}$ we find instead that the
  limits from RG-induced $K$ decays and tree-level $B$ decays can be
  comparable. So as to not clutter the plot, for these cases
  we show only the bounds from $B$ decays.
As can be seen from Fig.~\ref{fig:bounds}, in all cases where $ s $ or $ d $ are sequestered,
the limits remain rather weak, implying only $m_a \lesssim 1$ eV.

The axion-photon coupling is given by
$g_{a\gamma} = \frac{\alpha_\mathrm{em}}{2\pi f_a} [ E/N - 1.92(4) ]$
\cite{diCortona:2015ldu}
and depends on the ratio $E/N$ between the electromagnetic and QCD anomalies, which 
can acquire values within a rather large range, see~\cite{DiLuzio:2016sbl,DiLuzio:2017pfr}.
However, given that the leptons carry electromagnetic charge, the value of this ratio cannot be determined without first establishing how the $U(1)$ symmetry acts in the lepton sector.
A detailed exploration of the lepton sector is beyond the scope of this paper (see e.g.~\cite{DiLuzio:2017pfr} for a recent reassessment), but for illustration we consider the two
simplest scenarios, wherein all leptons couple either to $\tilde{H}_1$ or to $\tilde{H}_2$.
For each scenario we have up to four possible values of $ E/N $: 
since the up- and down-type quarks have different electric charges, 
their contribution to $ E $ depends on how they are assigned to $ M_{4,5} $, 
as well as on the choice of Yukawa structure $ \mathcal{T}_{1,2} $.
If leptons couple to $ \tilde{H}_1 $, $ E/N = 2/3 $, $ -1/3 $ or $ -10/3 $,
while if they couple to $ \tilde{H}_2 $, $ E/N = 8/3 $, $ 11/3 $ or $ 20/3 $.
As these values yield similar predictions for $g_{a\gamma}$, in Fig.~\ref{fig:bounds} 
we plot $ |g_{a\gamma}| $ only for the extremal cases, corresponding to $ E/N = 8/3 $ and $ 20/3 $.
The highlighted segments on these lines correspond to
$m_a \in [25,150]$ $\mu$eV, which is favored by the calculation of
the axion DM relic density in the post-inflationary PQ-breaking
scenarios.\footnote{%
A recent computation of the decay of topological defects
  and their contribution to the axion relic DM density in the
  post-inflation scenarios with $N_\mathrm{DW} = 1$ predicts a range $m_a \in [60,150]$ $\mu$eV
  \cite{Kawasaki:2014sqa,Borsanyi:2016ksw}. Another study claims a
  more definite and lower prediction $m_a = 26.5 \pm 3.4$ $\mu$eV
  \cite{Klaer:2017ond}, while Ref.~\cite{Gorghetto:2018myk} argues that 
  the uncertainty on the relic axion mass is much larger due to the required 
  extrapolation of the axion spectrum 
  from strings till the time of the QCD crossover. 
}
Observationally, $ g_{a\gamma} $ is constrained by measurements of
stellar cooling rates, and chiefly from the evolution of horizontal
branch (HB) stars in globular clusters \cite{Ayala:2014pea}.  Upper
bounds on $ |g_{a\gamma}| $ (depicted in Fig. \ref{fig:bounds} with full
lines) are also set by CAST \cite{Anastassopoulos:2017ftl} and ADMX
\cite{Du:2018uak} for different $ m_a $.  The sensitivities of
future axion experiments are shown with dashed lines.
Projections from IAXO \cite{Armengaud:2014gea} are given in blue,
while in green we have, from left to right, projections from
KLASH \cite{Alesini:2017ifp},
ACTION \cite{Choi:2017hjy},
ADMX,
CULTASK \cite{Chung:2016ysi}, and
% HEYSTAC \cite{Zhong:2018rsr}, 
MADMAX \cite{TheMADMAXWorkingGroup:2016hpc}.
For a recent review of axion experiments, see \cite{Irastorza:2018dyq}.
A separate upper bound on $ m_a $ is given by limits on hot DM
abundance from structure formation \cite{DiValentino:2015wba} which is
depicted in the figure with a vertical grey line.  

\subsection{Astrophobia}
\label{sec:nucleophobia}

It was recently observed \cite{DiLuzio:2017ogq} that in DFSZ-like
axion models with non-universal quark PQ charges, it is possible to
suppress the axion couplings to both protons ($C_p$) and neutrons
($C_n$), provided certain conditions are met for the axion coupling to
the lightest quarks.  This is important since, contrary to common
belief, it implies that the oft-quoted bound on the axion mass,
$m_a \lesssim 20$ meV, from the neutrino burst duration of the
Supernova SN1987A \cite{Tanabashi:2018oca,Raffelt:2006cw} can be
significantly relaxed, thus calling for exploration of the
$\mathcal{O}(0.1)$ eV axion mass region.

Conforming to standard notation, we define $C_u \equiv \cAx^u_{11}$
and $C_d \equiv \cAx^d_{11}$, respectively for the axion coupling to
up and down quarks. Following \cite{DiLuzio:2017ogq}, let us consider
the combinations $C_p \pm C_n$, given by
\begin{equation}
\begin{aligned}
    C_p + C_n &= 0.50(5) \left(C_u + C_d - 1 \right) - 2\delta_s , \\
    C_p - C_n &= 1.273(2) \left(C_u - C_d - \frac{1}{3} \right) ,
\end{aligned}
\end{equation}
where we used the relations $ f_u + f_d = 1 $ (exact) and
$ f_u - f_d \simeq 1/3 $ (approximate).  $|\delta_s| \lesssim 0.04$ is
a correction dominated by the strange sea quark
\cite{diCortona:2015ldu}.  The conditions for nucleophobia read
$C_p \pm C_n \approx 0$.  While the sum does not depend on the vev
angle $\beta$, the difference does, i.e.~there exist specific values
of $\tan\beta$ (see below) for which $C_p - C_n \approx 0$.  In the
class of models discussed here all charges and couplings are known,
and we find that, provided we sequester either the up or down quark,
the charge assignments yielding the $\mathcal{T}_2$ structure allow us to
enforce such a cancellation, i.e. nucleophobic models can be
constructed by choosing $\mathcal{T}_2^{(q)}$ with $ q=u,d $.  To
show that this is indeed the case, let us consider the limit where
weak and mass eigenstates coincide.
In this limit, up to independent reordering of their columns and concurrent
reordering of the rows, the Yukawa matrices $M_{4,5}$ are diagonal.
We can then extract the individual contributions from each generation
to the anomaly from the positions of the non-zero entries in the mass
matrices.  Let us consider the case of $ \mathcal{T}^{(d)}_2 $.  We
are still free to choose the ordering of the weak eigenstates with
respect to the mass basis, and we choose the usual one, introducing an
index $i=1,2,3$ that labels the mass eigenstates $ \hat{u}_{Li} $,
$ \hat{u}_{Ri} $, $ \hat{d}_{Li} $, and $ \hat{d}_{Ri} $ from light to
heavy.  As $ M_u $ is diagonal, and given the charges in
\eqn{eq:finalcharges}, we have
$ \mX(\hat{u}_{L1}) \equiv \mX(\hat{d}_{L1}) = \mX(Q)_1 = 2 $ and
$ \mX(\hat{u}_{R1}) = \mX(u)_1 = 3-c^2_\beta $, and so on for
$ i = 2, 3 $.  For $\mathcal{T}_2$, $ M_d $ is instead anti-diagonal,
so we have
$ \mX(\hat{d}_{R1}) = \mX(d)_3 = 2+c^2_\beta $,
$ \mX(\hat{d}_{R2}) = \mX(d)_2 = c^2_\beta $, and
$ \mX(\hat{d}_{R3}) = \mX(d)_1 = -1+c^2_\beta $.  
The contribution of the single generations to the anomaly 
then are $ 2N_1 = 1 $ and $ 2N_2 = 2N_3 = 0 $.  (For $\mathcal{T}_1 $
we would instead obtain $ 2N_1 = 2N_2 = 1 $ and $2N_3 = 0 $.)  The
first nucleophobia condition $ C_u + C_d = 1 $ is realized when
$ N_l/N = 1 $, where $N_l$ is the anomaly contribution of the lightest
generation. This implies that the contributions of the two heavier
generations must add up to zero.  This can be satisfied for
$\mathcal{T}_2 $ (but not for $\mathcal{T}_1 $).  It is sufficient to
arrange for $N_l = N_1$ which can be done by sequestering one
of the two light quarks.  Taking into account the effects of quark
mixing, we obtain $C_u + C_d \simeq 1.1$ which might even result in a
more accurate cancellation also against the $\delta_s$ correction.
Barring fine-tunings between mixings and strange quark corrections, we
see that just from the choice of a specific model, and without any
tuning of the parameters, we can obtain a suppression of the
combination $C_p + C_n$ at the level of 10\%.  The second condition
for nucleophobia is satisfied for $C_u - C_d \approx 1/3$. This can be
obtained by an appropriate choice for the ratio of Higgs doublet
vevs.  We need $\tan \beta \approx 1/\sqrt{2}$ for
$\mathcal{T}_2^{(u)}$ , and $\tan \beta \approx \sqrt{2}$ for
$\mathcal{T}_2^{(d)}$, tuned to within about 4\%,
in order to match the accuracy of the first condition.

For standard DFSZ models, in which the leptons and in particular the
electrons are also coupled to the axion, bounds from white dwarf
cooling rates and from red giant evolution apply, and yield limits on
the axion mass which are not much weaker than those from
SN1987A~\cite{Tanabashi:2018oca}.  However, in some models also the
axion-electron coupling can be suppressed, thus making the axion
``astrophobic''~\cite{DiLuzio:2017ogq}, in the sense that all the
strongest bounds from astrophysical considerations can be relaxed.
There are at least two ways to achieve electrophobia. In the first
case, the Higgs sector is not enlarged, and the leptons couple to the
same Higgs doublets as do quarks. The electrons carry a
PQ charge and thus couple to the axion. However, the lepton
sector is characterized by large flavour mixings and, as was shown
in \cite{DiLuzio:2017ogq}, by an appropriate choice of lepton $U(1)$
charges and Yukawa matrix structures one can enforce a cancellation
between the contribution of the unmixed couplings and large
corrections from mixing, while maintaining agreement with the
lepton sector experimental data.
The second approach is to couple all leptons exclusively to a third Higgs doublet $ H_3 $ with charge $ \mX_3 $, 
which in turn couples to $ H_{1,2} $ via judiciously chosen non-Hermitian invariants in the scalar potential.
It can be shown that after imposing orthogonality of the axion with respect to the hypercharge Goldstone mode,
it is possible to arrange the ratios of the vevs such that the redefined $ \mX_3 $ charge becomes arbitrarily small, effectively decoupling the whole lepton sector from the physical axion.

\section{Conclusion}
\label{sec:conclusions}

In this paper we have explored a class of models for the quark Yukawa
sector which are characterized by having the maximum possible reduction in
the number of fundamental Yukawa operators, while still ensuring
compatibility with all experimental data.  This demand for minimality
can be motivated by Occam’s Razor: to explain the same set of
data, a generic SM Yukawa structure involves twice as many Yukawa
operators as we have assumed here, and therefore our Yukawa structures are simpler by far. 
To comply with another well-known paradigm, namely
Gell-Mann's totalitarian principle stating ``everything not forbidden
is compulsory'', we have appealed to a simple, generation-dependent
$U(1)$ symmetry, that forbids in total nine Yukawa operators, and
generates corresponding texture zeros in the Yukawa matrices, at the cost of introducing a second Higgs doublet.  
We have shown that there are only two Yukawa
structures, inequivalent under row and column permutations, that can
ensure all quark masses, CKM mixings and CP phase are non-zero, while also corresponding to a consistent set of $U(1)$
charge assignments for the quarks and Higgs fields.
Of the nine non-zero Yukawa couplings, all but one can be made real by
quark redefinitions, and we have provided a method to determine
univocally the numerical values of these 9+1 input parameters in terms
of experimental observables.  We have found that the
generation-dependent $U(1)$ symmetries which enforce these minimal
Yukawa structures all have a QCD anomaly, and thus can be
straightforwardly interpreted as PQ symmetries.  The resulting axion
is characterized by couplings that feel the flavour content of the
quarks to which it couples, mediating flavour-changing
meson decays that may be visible in future experiments.

Interestingly, two specific Yukawa textures allow for the construction of models
wherein the axion couplings to nucleons can be suppressed by one
order of magnitude, with a moderate amount of tuning in the parameters.
This gives rise to a nucleophobic axion, for which the strong
constraints from the neutrino burst duration of the SN1987A are relaxed. 
We have also described two different ways to
suppress the axion couplings to electrons. The first one does not
require enlarging the Higgs sector, but it relies on a tuned
cancellation in the axion-electron coupling. The second one requires a
third Higgs doublet, but with no need for additional tunings in the
parameters beyond what is already required to enforce
nucleophobia.  Such axions truly deserve the title of
``astrophobic''~\cite{DiLuzio:2017ogq}, since for all of them the
strongest astrophysical bounds are sizeably relaxed, which renders a region of relatively large axion masses, $0.1 \lesssim m_a/\mathrm{eV} \lesssim 1$, generally believed to be ruled out, indeed viable.
Finally, it would be interesting to extend the present study to the lepton sector, where the scale of $ U(1) $ breaking could be naturally connected with the seesaw scale.

\section*{Acknowledgments}
We acknowledge valuable discussions with Paolo Panci and Robert Ziegler in the early stages of this work.
FB and EN are supported in part by the INFN ``Iniziativa Specifica'' Theoretical Astroparticle Physics (TAsP-LNF).
The work of LDL was supported by the ERC grant NEO-NAT.
This work is supported by MINECO grant FPA2016-76005-C2-1-P, by Maria de Maetzu program grant MDM-2014-0367 of ICCUB and 2017 SGR 929.
The work of EN was performed in part at the Aspen Center for Physics, which is supported by National Science Foundation grant PHY-1607611. 
The participation of EN at the Aspen Center for Physics was supported by the Simons Foundation. 

\appendix

\section{3+6 textures}
\label{app:3p6}

In this appendix, we show that the $ M_3 \oplus M_6 $ Yukawa structures are not compatible with a consistent $ U(1) $ charge assignment of the quarks.
Let us choose $ M_d = M_3 $, with three non-zero Yukawa textures, and $ M_u = M_6 $.
By appropriate choice of $ U(1) $ charges $ \mX(d) $, we arrange the only diagonal elements of $ \mX_{\Qbar d} $ to be $ \pm s $, where $ s = \pm 1 $ correspond to the two Higgs charges.
All off-diagonal elements must be $ \neq \pm s $, leading to constraints on the quark $ U(1) $ charges.
There are two possibilities for the diagonal entries of $ \mX_{\Qbar d} $, namely all diagonal charges equal, i.e.~$ (s,s,s) $ or one charge different, i.e.~$ (s,s,-s) $ or some permutation thereof.
Denoting the row charge difference between two rows $ i $ and $ j $ as $ \Delta^Q_{ij} $, we have
\begin{equation}
	\mX_{\Qbar d} = \pmatr{
		\bs & s+\Delta^Q_{12} & \pm s+\Delta^Q_{13} \\
		s-\Delta^Q_{12} & \bs & \pm s+\Delta^Q_{23} \\
		s-\Delta^Q_{13} & s-\Delta^Q_{23} & \pm \bs
	}.
\end{equation}
Regardless of the sign in the third column, we thus require $ \Delta^Q_{12} \neq 0, \pm 2s $, $ \Delta^Q_{23} \neq 0, 2s $, and $ \Delta^Q_{13} \neq 0, 2s $.
Additionally if $ \diag(\mX_{\Qbar d}) = (s,s,s) $, we also have $ \Delta^Q_{23} \neq -2s $ and $ \Delta^Q_{13} \neq -2s $.

We will now see that these constraints exclude all physically viable textures in $ M_u = M_6 $. 
We must fill six entries; it is immediately clear that at least two columns of $ M_6 $ will have at least two filled entries each.
In fact, given that each row in $ M_3 $ (and thereby also in $ M_6 $) is different, each column has exactly two non-zero entries.
In a given column, in order to have non-zero Yukawa couplings, the bilinear $ U(1) $ charges must be $ \pm s $ and the difference between them, i.e.~$ \mX_{ik} - \mX_{jk} $, must be $ 0 $ or $ \pm 2s $.
Most of these choices are disallowed by the constraints on $ M_3 $.
The only possible options are $ \Delta^Q_{13} = -2s $ and $ \Delta^Q_{23} = -2s $.
Choosing $ \Delta^Q_{23} = -2s $, we can arrange two identical columns in a charge consistent way, like
\begin{equation}
	\mX_{\Qbar u} = \pmatr{
		ks & ks & \cdot \\
		-\bs & -\bs & \cdot \\
		\bs & \bs & \cdot  
	}_{k \neq \pm 1}.
\end{equation}
In order to ensure a mass matrix with a non-zero determinant, i.e.~no rows or columns of zeroes in $ M_6 $, the final column must take the form $ (\pm \bs, \bs, k^\prime s) $ or $ (\pm \bs, k^\prime s, -\bs) $, where $ k^\prime \neq \pm 1 $. 
However, any of these configurations violates one of the constraints on $ \Delta^Q_{12} $ or $ \Delta^Q_{13} $.
We reach the same conclusion if we choose $ \Delta^Q_{13} = -2s $.

Physically speaking, the reason 3+6 textures are excluded lies in the fact that enforcing a strictly diagonal $ M_3 $ texture places severe constraints on the allowed charges of the LH quarks.
The constraints in turn are not compatible with a much richer mass matrix structure in $ M_6 $.

\section{Numerical fits}
\label{app:fits}

Recall that there are two structures, $\mathcal{T}_1$ and $\mathcal{T}_2$, given in \eqn{eq:textures}.
For each, we add a superscript $(q)$, where $q = u$, $c$, $t$, $d$, $s$, $b$ labels the quark in $M_4$ that is ``sequestered''.
The model parameters can be obtained by the method outlined in Section \ref{sec:fit}.
The vector ($\cVec^q$) and axial ($\cAx^q$) coupling matrices of the axion to quarks ($q = u, d$ labels the sectors) are defined in \eqn{eq:VfAf}.
Notably, the sequestered quark has no FV couplings.
Table \ref{tab:fitandfvcouplings} presents the model parameters corresponding to the correct quark masses at $\mu = 4 \times 10^{12}$ GeV, which is close to the presumed $U(1)$-breaking scale.
It also gives the off-diagonal elements of $\cVec^q$ and $\cAx^q$, which describe flavour-violating interactions.

\begin{table}[ht]
\newcommand{\bmatr}[1]{\begin{bmatrix}#1\end{bmatrix}}
\centering
\scriptsize
\setlength\arraycolsep{2pt}
\begin{tabular}{c@{\hskip 5pt}cc crrrr}
\toprule
	Texture &$ M_u \,/ $GeV & $ M_d \,/ $GeV& $(i,j)$ & $ \cVec^u_{ij} $ & $ \cAx^u_{ij} $ & $ \cVec^d_{ij} $ & $ \cAx^d_{ij} $ \\
\midrule
	\multirow{3}{*}{$ \mathcal{T}_1^{(u)} $} &
	\multirow{3}{*}{$\bmatr{0.28 & 4.9 e^{i \rho_u} & 0 \\ 0 & 83 & 0 \\ 0 & 0 & 0.00061}$} & 
	\multirow{3}{*}{$\bmatr{0.0055 & 0.043 & 0 \\ 0 & 0.95 & 0.67 \\ 0 & 0 & 0.0073}$} & 
	(1,2) & $ 0 $ & $ 0 $ & $ 0.39 $ & $ -0.051 $ \\
&&&	(1,3) & $ 0 $ & $ 0 $ & $ 0.047 $ & $ 0.05 $ \\
&&&	(2,3) & $ -0.03 $ & $ 0.03 $ & $ -0.25 $ & $ -0.22 $ \\[1.5ex]
	\multirow{3}{*}{$ \mathcal{T}_1^{(c)} $} &
	\multirow{3}{*}{$\bmatr{0.00061 & 0.43 e^{i \rho_c} & 0 \\ 0 & 83 & 0 \\ 0 & 0 & 0.28}$} &
	\multirow{3}{*}{$\bmatr{0.0013 & 0.0066 & 0 \\ 0 & 0.54 & 1.0 \\ 0 & 0 & 0.055}$} & 
	(1,2) & $ 0 $ & $ 0 $ & $ 0.23 $ & $ -0.21 $ \\
&&&	(1,3) & $ -0.003 $ & $ 0.003 $ & $ 0.006 $ & $ -0.001 $ \\
&&&	(2,3) & $ 0 $ & $ 0 $ & $ -0.23 $ & $ -0.18 $ \\[1.5ex]
	\multirow{3}{*}{$ \mathcal{T}_1^{(t)} $} &
	\multirow{3}{*}{$\bmatr{0.00061 & 0.024 e^{i \rho_t} & 0 \\ 0 & 0.28 & 0 \\ 0 & 0 & 83}$} &
	\multirow{3}{*}{$\bmatr{0.0013 & 0.0055 & 0 \\ 0 & 0.025 & 0.049 \\ 0 & 0 & 1.2}$} & 
	(1,2) & $ -0.043 $ & $ 0.043 $ & $ -0.11 $ & $ 0.098 $ \\
&&&	(1,3) & $ 0 $ & $ 0 $ & $ 0.004 $ & $ -0.004 $ \\
&&&	(2,3) & $ 0 $ & $ 0 $ & $ 0.021 $ & $ -0.020 $ \\[1.5ex]
	\multirow{3}{*}{$ \mathcal{T}_1^{(d)} $} &
	\multirow{3}{*}{$\bmatr{0.0027 & 0.28 & 0 \\ 0 & 7.1 & 82 \\ 0 & 0 & 0.74}$} &
	\multirow{3}{*}{$\bmatr{0.026 & 0.048 e^{i \rho_d} & 0 \\ 0 & 1.2 & 0 \\ 0 & 0 & 0.0013}$} & 
	(1,2) & $ 0.22 $ & $ -0.22 $ & $ 0 $ & $ 0 $ \\
&&&	(1,3) & $ 0.005 $ & $ -0.004 $ & $ 0 $ & $ 0 $ \\
&&&	(2,3) & $ -0.044 $ & $ -0.042 $ & $ -0.021 $ & $ 0.020 $ \\[1.5ex]
	\multirow{3}{*}{$ \mathcal{T}_1^{(s)} $} &
	\multirow{3}{*}{$\bmatr{0.00063 & 0.064 & 0 \\ 0 & 6.6 & 82 \\ 0 & 0 & 3.4}$} &
	\multirow{3}{*}{$\bmatr{0.0013 & 0.010 e^{i \rho_s} & 0 \\ 0 & 1.2 & 0 \\ 0 & 0 & 0.026}$} & 
	(1,2) & $ 0.22 $ & $ -0.219 $ & $ 0 $ & $ 0 $ \\
&&&	(1,3) & $ 0.005 $ & $ -0.005 $ & $ -0.004 $ & $ 0.004 $ \\
&&&	(2,3) & $ -0.06 $ & $ -0.020 $ & $ 0 $ & $ 0 $ \\[1.5ex]
	\multirow{3}{*}{$ \mathcal{T}_1^{(b)} $} &
	\multirow{3}{*}{$\bmatr{0.00061 & 0.024 & 0 \\ 0 & 0.28 & 3.5 \\ 0 & 0 & 83}$} &
	\multirow{3}{*}{$\bmatr{0.0013 & 0.0055 e^{i \rho_b} & 0 \\ 0 & 0.025 & 0 \\ 0 & 0 & 1.2 }$ }& 
	(1,2) & $ -0.043 $ & $ 0.043 $ & $ -0.11 $ & $ 0.097 $ \\
&&&	(1,3) & $ -0.002 $ & $ 0.002 $ & $ 0 $ & $ 0 $ \\
&&&	(2,3) & $ -0.021 $ & $ 0.021 $ & $ 0 $ & $ 0 $ \\
\midrule 
	\multirow{3}{*}{$ \mathcal{T}_2^{(u)} $} &
	\multirow{3}{*}{$\bmatr{0 & 0 & 0.00061\\ 0 & 83 & 0 \\ 0.28 & 4.9 e^{i \rho_u} & 0 }$} &
	\multirow{3}{*}{$\bmatr{0.0013 & 0.0072 & 0\\ 0& 0.68 & 0.94 \\ 0& 0 & 0.043  }$ }& 
	(1,2) & $ 0 $ & $ 0 $ & $ 0.45 $ & $ -0.424 $ \\
&&&	(1,3) & $ 0 $ & $ 0 $ & $ 0.009 $ & $ 0.002 $ \\
&&&	(2,3) & $ -0.06 $ & $ 0.06 $ & $ -0.51 $ & $ -0.44 $ \\[1.5ex]
	\multirow{3}{*}{$ \mathcal{T}_2^{(c)} $} &
	\multirow{3}{*}{$\bmatr{0 & 0 & 0.28\\ 0 & 83 & 0 \\ 0.00061 & 0.43 e^{i \rho_c} & 0 }$} &
	\multirow{3}{*}{$\bmatr{0.0055 & 0.055 & 0 \\ 0& 1.0 & 0.52  \\ 0 & 0 & 0.0067 }$ }& 
	(1,2) & $ 0 $ & $ 0 $ & $ 0.8 $ & $ -0.076 $ \\
&&&	(1,3) & $ -0.005 $ & $ 0.005 $ & $ 0.076 $ & $ 0.09 $ \\
&&&	(2,3) & $ 0 $ & $ 0 $ & $ -0.44 $ & $ -0.35 $ \\[1.5ex]
	\multirow{3}{*}{$ \mathcal{T}_2^{(t)} $} &
	\multirow{3}{*}{$\bmatr{0 & 0 & 83 \\ 0 & 0.28 & 0 \\ 0.00061 & 0.024 e^{i \rho_t} & 0 }$} &
	\multirow{3}{*}{$\bmatr{0.139 & 1.15 &0 \\ 0 & 0.049 & 0.025 \\0 & 0 & 0.0056 }$ }& 
	(1,2) & $ -0.086 $ & $ 0.086 $ & $ 0.64 $ & $ 0.22 $ \\
&&&	(1,3) & $ 0 $ & $ 0 $ & $ -0.12 $ & $ -0.11 $ \\
&&&	(2,3) & $ 0 $ & $ 0 $ & $ -0.068 $ & $ 0.014 $ \\[1.5ex]
	\multirow{3}{*}{$ \mathcal{T}_2^{(d)} $} &
	\multirow{3}{*}{$\bmatr{0.00063 & 0.74 & 0 \\ 0& 82 & 7.1  \\ 0& 0 & 0.28  }$} &
	\multirow{3}{*}{$\bmatr{0 & 0 & 0.0013\\ 0 & 1.2 & 0 \\ 0.026 & 0.048 e^{i \rho_d} & 0 }$ }& 
	(1,2) & $ 0.44 $ & $ -0.44 $ & $ 0 $ & $ 0 $ \\
&&&	(1,3) & $ -0.009 $ & $ 0.009 $ & $ 0 $ & $ 0 $ \\
&&&	(2,3) & $ -0.088 $ & $ -0.084 $ & $ -0.042 $ & $ 0.041 $ \\[1.5ex]
	\multirow{3}{*}{$ \mathcal{T}_2^{(s)} $} &
	\multirow{3}{*}{$\bmatr{0.0027 & 3.4 & 0 \\ 0 & 82 & 6.6 \\ 0 & 0 & 0.064 }$} &
	\multirow{3}{*}{$\bmatr{0 & 0 & 0.026\\ 0 & 1.2 & 0 \\  0.0013 & 0.010 e^{i \rho_s} & 0 }$ }& 
	(1,2) & $ 0.46 $ & $ -0.42 $ & $ 0 $ & $ 0 $ \\
&&&	(1,3) & $ -0.008 $ & $ 0.01 $ & $ -0.009 $ & $ 0.009 $ \\
&&&	(2,3) & $ -0.12 $ & $ -0.039 $ & $ 0 $ & $ 0 $ \\[1.5ex]
	\multirow{3}{*}{$ \mathcal{T}_2^{(b)} $} &
	\multirow{3}{*}{$\bmatr{0.17 & 83 & 0\\ 0& 3.5 & 0.28  \\ 0 & 0 & 0.024 }$} &
	\multirow{3}{*}{$\bmatr{0 & 0 & 1.2\\ 0 & 0.025 & 0 \\  0.0013 & 0.0055 e^{i \rho_b} & 0 }$ }& 
	(1,2) & $ 0.14 $ & $ -0.036 $ & $ -0.22 $ & $ 0.19 $ \\
&&&	(1,3) & $ -0.006 $ & $ 0.002 $ & $ 0 $ & $ 0 $ \\
&&&	(2,3) & $ -0.042 $ & $ 0.042 $ & $ 0 $ & $ 0 $ \\
\bottomrule
\end{tabular}
\caption{Input parameters $ m^q_{ij} $ fitted to flavour data at $ \mu = 4 \times 10^{12} $ GeV, and associated off-diagonal (flavour-violating) axion-quark couplings $\cVec^q_{ij} = \cVec^q_{ji}$ and $\cAx^q_{ij} = \cAx^q_{ji}$.
The phases $ \rho_i $ are as follows: $ \rho_u = 0.7179 $, $ \rho_c = -0.7162 $, $ 
\rho_t = 1.5681 $, $ \rho_d = -0.3994 $, $ \rho_s = 0.3988 $, and $ \rho_b = -1.5693 $.
}
\label{tab:fitandfvcouplings}
\end{table}

Table \ref{tab:fdiagcouplings} gives the corresponding diagonal elements of the coupling matrices, which depend explicitly on $\beta = \arctan (v_2/v_1)$.
Note that the couplings are all real: this follows from the highly constrained phase structure of the Yukawa matrices, with only one non-zero phase in the off-diagonal element of $M_4$, which ultimately cancels in $\cVec^q$ and $\cAx^q$.

\begin{table}[ht]
\newcommand{\cosb}{c^2_\beta}
\newcommand{\exto}[1]{\cdot\!10^{#1}}
\centering
\scriptsize
\setlength\arraycolsep{2pt}
\begin{tabular}{cc rr rr}
\toprule
  Texture & $i$ & $ \cVec^u_{ii} $ & $ \cAx^u_{ii} $ & $ \cVec^d_{ii} $ & $ \cAx^d_{ii} $ \\
\midrule
\multirow{3}{*}{$ \mathcal{T}_1^{(u)} $} 
& 1 & $0.0  -\cosb/2   $ & $0.0  -\cosb/2   $ & $1.0 +\cosb/2   $ & $0.91 +\cosb/2   $ \\ 
& 2 & $2.5  -\cosb/2   $ & $0.5  -\cosb/2   $ & $1.2 +\cosb/2   $ & $-0.75 +\cosb/2   $ \\ 
& 3 & $1.5  -\cosb/2   $ & $0.5  -\cosb/2   $ & $0.83 +\cosb/2   $ & $-0.17 +\cosb/2   $ \\[1.5ex] 
\multirow{3}{*}{$ \mathcal{T}_1^{(c)} $} 
& 1 & $2.5  -\cosb/2   $ & $0.5  -\cosb/2   $ & $1.9 +\cosb/2   $ & $0.051 +\cosb/2   $ \\ 
& 2 & $0.0  -\cosb/2   $ & $0.0  -\cosb/2   $ & $0.45 +\cosb/2   $ & $0.34 +\cosb/2   $ \\ 
& 3 & $1.5  -\cosb/2   $ & $0.5  -\cosb/2   $ & $0.61 +\cosb/2   $ & $-0.39 +\cosb/2   $ \\[1.5ex] 
\multirow{3}{*}{$ \mathcal{T}_1^{(t)} $} 
& 1 & $2.5  -\cosb/2   $ & $0.5  -\cosb/2   $ & $2.0 +\cosb/2   $ & $0.022 +\cosb/2   $ \\ 
& 2 & $1.5  -\cosb/2   $ & $0.5  -\cosb/2   $ & $1.0 +\cosb/2   $ & $-0.021 +\cosb/2   $ \\ 
& 3 & $0.0  -\cosb/2   $ & $0.0  -\cosb/2   $ & $0.0009 +\cosb/2   $ & $-0.0009 +\cosb/2   $ \\[1.5ex] 
\multirow{3}{*}{$ \mathcal{T}_1^{(d)} $} 
& 1 & $1.6 -  \cosb/2$ & $1.4 -  \cosb/2$ & $-0.5+ \cosb/2$ & $-0.5+ \cosb/2$ \\ 
& 2 & $1.9 -  \cosb/2$ & $0.047 -  \cosb/2$ & $2.0+ \cosb/2$ & $8.6\exto{-4}+ \cosb/2$ \\ 
& 3 & $1.0 -  \cosb/2$ & $0.0037 -  \cosb/2$ & $1.0+ \cosb/2$ & $-8.6\exto{-4}+ \cosb/2$ \\[1.5ex] 
\multirow{3}{*}{$ \mathcal{T}_1^{(s)} $} 
& 1 & $2.4 -  \cosb/2$ & $0.55 -  \cosb/2$ & $2.0+ \cosb/2$ & $4.0\exto{-5}+ \cosb/2$ \\ 
& 2 & $1.0 -  \cosb/2$ & $0.95 -  \cosb/2$ & $-0.5+ \cosb/2$ & $-0.5+ \cosb/2$ \\ 
& 3 & $1.0 -  \cosb/2$ & $0.004 -  \cosb/2$ & $1.0+ \cosb/2$ & $-4.0\exto{-5}+ \cosb/2$ \\[1.5ex] 
\multirow{3}{*}{$ \mathcal{T}_1^{(b)} $} 
& 1 & $2.5 -  \cosb/2$ & $0.5 -  \cosb/2$ & $2.0+ \cosb/2$ & $0.022+ \cosb/2$ \\ 
& 2 & $1.5 -  \cosb/2$ & $0.5 -  \cosb/2$ & $1.0+ \cosb/2$ & $-0.022+ \cosb/2$ \\ 
& 3 & $0.5 -  \cosb/2$ & $0.5 -  \cosb/2$ & $-0.5+ \cosb/2$ & $-0.5+ \cosb/2$ \\ 
\midrule 
% \rule{0pt}{3ex}
\multirow{3}{*}{$ \mathcal{T}_2^{(u)} $} 
& 1 & $5.0   -\cosb   $ & $ 1.0 -\cosb$ & $3.9 +\cosb   $ & $0.1 +\cosb   $ \\ 
& 2 & $0.0036  -\cosb   $ & $-0.0036  -\cosb   $ & $0.76 +\cosb   $ & $0.55 +\cosb   $ \\ 
& 3 & $2.0  -\cosb   $ & $0.0036  -\cosb   $ & $1.3 +\cosb   $ & $-0.66 +\cosb   $ \\[1.5ex] 
\multirow{3}{*}{$ \mathcal{T}_2^{(c)} $} 
& 1 & $0.00003  -\cosb   $ & $-0.00003 -\cosb   $ & $2.0 +\cosb   $ & $1.8 +\cosb   $ \\ 
& 2 & $5.0   -\cosb   $ & $1 -\cosb$ & $2.2 +\cosb   $ & $-1.6 +\cosb   $ \\ 
& 3 & $2.0  -\cosb   $ & $0.00003  -\cosb   $ & $1.8 +\cosb   $ & $-0.21 +\cosb   $ \\[1.5ex] 
\multirow{3}{*}{$ \mathcal{T}_2^{(t)} $} 
& 1 & $0.0074 -\cosb$ & $-0.0074 -\cosb$ & $1.9 +\cosb   $ & $1.8 +\cosb   $ \\ 
& 2 & $2.0 -\cosb$ & $0.0074 -\cosb$ & $1.1 +\cosb   $ & $-0.86 +\cosb   $ \\
& 3 & $5.0 -\cosb$ & $ 1.0 -\cosb$ & $3.0 +\cosb   $ & $-0.98 +\cosb   $ \\[1.5ex]  
\multirow{3}{*}{$ \mathcal{T}_2^{(d)} $} 
& 1 & $4.9 -  \cosb$ & $1.1 -  \cosb$ & $4.0 + \cosb$ & $ \cosb$ \\ 
& 2 & $1.1 -  \cosb$ & $0.91 -  \cosb$ & $-1.0+ \cosb$ & $-1.0+ \cosb$ \\ 
& 3 & $3.0 -  \cosb$ & $0.99 -  \cosb$ & $1.0+ \cosb$ & $-1.0+ \cosb$ \\[1.5ex] 
\multirow{3}{*}{$ \mathcal{T}_2^{(s)} $} 
& 1 & $3.1 -  \cosb$ & $2.9 -  \cosb$ & $-1.0+ \cosb$ & $-1.0+ \cosb$ \\ 
& 2 & $2.9 -  \cosb$ & $-0.89 -  \cosb$ & $4.0 + \cosb$ & $ \cosb$ \\ 
& 3 & $3.0 -  \cosb$ & $0.99 -  \cosb$ & $1.0+ \cosb$ & $-1.0+ \cosb$ \\[1.5ex]
\multirow{3}{*}{$ \mathcal{T}_2^{(b)} $} 
& 1 & $3.0 -  \cosb$ & $3.0 -  \cosb$ & $-0.96+ \cosb$ & $-1.0+ \cosb$ \\ 
& 2 & $2.0 -  \cosb$ & $0.0069 -  \cosb$ & $0.96+ \cosb$ & $-0.96+ \cosb$ \\ 
& 3 & $4.0 -  \cosb$ & $0.0018 -  \cosb$ & $4.0 + \cosb$ & $ \cosb$ \\ 
\bottomrule
\end{tabular}
\caption{Diagonal vector ($\cVec^q_{ii}$) and axial-vector ($\cAx^q_{ii}$) couplings of the axion to quarks. 
}
\label{tab:fdiagcouplings}
\end{table}

\clearpage

\providecommand{\href}[2]{#2}\begingroup\raggedright\endgroup

\end{document}